\documentclass[twocolumn,aps,prl,showpacs,superscriptaddress,unsortedaddress]{revtex4}
\usepackage{epsf}
\usepackage{graphicx}
\newcommand{\etal}{{\it et al.}}
\begin{document}

\title{Dynamic Response Functions from Angle Resolved Photoemission Spectra}

\author{U. Chatterjee}
\affiliation{Department of Physics, University of Illinois at Chicago, Chicago, IL 60607}
\affiliation{Materials Science Division, Argonne National Laboratory, Argonne, IL 60439}
\author{D. K. Morr}
\affiliation{Department of Physics, University of Illinois at Chicago, Chicago, IL 60607}
\author{M. R. Norman}
\affiliation{Materials Science Division, Argonne National Laboratory, Argonne, IL 60439}
\author{M. Randeria}
\affiliation{Department of Physics, The Ohio State University, Columbus, OH  43210}
\author{A. Kanigel}
\affiliation{Department of Physics, University of Illinois at Chicago, Chicago, IL 60607}
\author{M. Shi}
\affiliation{Department of Physics, University of Illinois at Chicago, Chicago, IL 60607}
\affiliation{Swiss Light Source, Paul Scherrer Institut, CH-5232 Villigen, Switzerland}
\author{E. Rossi}
\affiliation{Department of Physics, University of Illinois at Chicago, Chicago, IL 60607}
\author{A. Kaminski}
\author{H. M. Fretwell}
\affiliation{Ames Laboratory and Department of Physics and Astronomy, Iowa State University,
Ames, IA 50011}
\author{S. Rosenkranz}
\affiliation{Materials Science Division, Argonne National Laboratory, Argonne, IL 60439}
\author{K. Kadowaki}
\affiliation{Institute of Materials Science, University of Tsukuba, Ibaraki 305-3573, Japan}
\author{J. C. Campuzano}
\affiliation{Department of Physics, University of Illinois at Chicago, Chicago, IL 60607}
\affiliation{Materials Science Division, Argonne National Laboratory, Argonne, IL 60439}

\date{\today}

\begin{abstract}
We introduce a formalism for calculating dynamic response functions using experimental
single-particle Green's functions.
As an illustration of this procedure we estimate the dynamic spin response of the cuprate
superconductor Bi$_2$Sr$_2$CaCu$_2$O$_{8+\delta}$. We find good agreement with superconducting state
neutron data, in particular the $(\pi,\pi)$ resonance with its unusual `hourglass' shaped
dispersion. We anticipate our formalism will also be useful in
interpreting results from other spectroscopies, such as optical and Raman responses.
\end{abstract}

\maketitle

The linear response to an external probe as a function of  momentum and frequency
is of great importance in elucidating the properties of complex materials.  Examples include various two-particle correlation functions involving spin, current and charge as
measured by inelastic neutron scattering (INS), nuclear magnetic resonance (NMR), optical conductivity, and Raman scattering experiments.  On the other hand, angle resolved photoemission spectroscopy (ARPES) \cite{REVIEW} directly gives information about single-particle excitations of a system.
The response function of a system
can be expressed in terms of a two-particle correlation function of the observable to which the external probe couples.
The goal of this paper
is to develop an approach to use single-particle spectroscopy data to gain insight into two-particle
correlation functions.  In particular we focus here on using the Green's functions
obtained from superconducting state  ARPES
data in the high T$_c$ cuprates to compute the dynamic spin
susceptibility, which we then compare with INS data \cite{NEUTRON}.

 From a theoretical point of view,
dynamic response functions are difficult to calculate in general and
many different approximate formalisms exist in the literature. For
instance, there are two rather different approaches for computing
the dynamic spin response for the high T$_c$ cuprate
superconductors. The first is based on the random phase
approximation (RPA) \cite{RPA} and related diagrammatic formulations
\cite{OTHERS}. This approach not only assumes momentum is a good
quantum number, but also that the spin and charge degrees of freedom
are coupled. The second is based on spin ladders separated by
one-dimensional domain walls known  as stripes.  In this formalism
spatial inhomogeneity is important, and the charge sector is assumed
to be secondary when calculating the spin response \cite{UHRIG}.
Despite the quite different physics underlying these two schemes,
the results for the calculated spin response function of the
cuprates are similar  - one of the current dilemmas facing the field
of high T$_c$ superconductivity. It is thus important to go beyond a
purely theoretical approach and directly employ information obtained
from one experiment (ARPES) to make progress on interpreting the
dynamic susceptibility measured by another (INS).

We use a formalism based on a diagrammatic ${\bf k}$-space approach
which goes beyond RPA in that it uses fully dressed Green's function
obtained from ARPES data on Bi$_2$Sr$_2$CaCu$_2$O$_{8+\delta}$
(Bi2212). We compare the calculated superconducting state
susceptibility with INS data. We obtain the $(\pi,\pi)$ resonance
seen in many cuprates \cite{NEUTRON}, including Bi2212 \cite{INSB},
and also its unusual `hourglass' shaped dispersion as observed in
YBa$_2$Cu$_3$O$_{7-\delta}$ (YBCO) \cite{ARAI,HAYDEN} and more
recently in Bi2212 \cite{FAUQUE}. We also find that the magnetic
dispersion is sensitive to the momentum dependence of the effective
interaction used to calculate the susceptibility.

We use ARPES spectra from a near-optimal sample (T$_c$=90K) of
Bi2212,  the data having been presented previously
\cite{KINK,UTPAL}. While a resonance peak was observed in this
material some time ago \cite{INSB}, a more detailed study with
results similar to the much more extensive INS data for YBCO, has
appeared only recently \cite{FAUQUE}.

Quite generally, two-particle correlation functions can be written
in terms of single-particle Green's functions and vertex parts
\cite{MAHAN}.
The lowest order term contributing to the spin susceptibility (the bare polarization bubble) in the superconducting state can be written as \cite{BOB}
\begin{widetext}
\begin{equation}
\chi_0({\bf q},\Omega)=\frac{1}{\pi^2} \sum_{\bf k}
\int_{-\infty}^\infty d\nu \ d\epsilon  \left[
{\rm Im}G({\bf k},\nu) \ {\rm Im}G({\bf k+q},\epsilon) + {\rm
Im}F({\bf k},\nu) \ {\rm Im}F({\bf k+q},\epsilon) \right]  \
\frac{n_F(\nu)-n_F(\epsilon)}{\Omega+\nu-\epsilon+i\delta}=\chi_0^G
+ \chi_0^F
\label{eq:chi0}
\end{equation}
\end{widetext}
where ${\rm Im}$ denotes the imaginary part of the normal  and
anomalous Green's functions $G$ and $F$, and $\chi_0^G$ and
$\chi_0^F$ denote the $GG$ and $FF$ contributions to $\chi_0$
respectively.

We next describe in detail how ${\rm Im}G$ is extracted from ARPES
data and return later to the question of estimating the contribution
of ${\rm Im}F$ (which is not directly measured). ARPES probes the
occupied part of the spectral function leading to the intensity
$I({\bf k},\omega) \propto n_F(\omega) \ {\rm Im}G({\bf k},\omega)$,
where $n_F(\omega)$ is the Fermi function \cite{NK}. In order to
extract ${\rm Im}G$ from raw data we need to address several issues
including data normalization, background subtraction, and removing
the effects of the Fermi function.  In addition we need to extend
${\rm Im}G$ to $\omega >0$ to calculate $\chi_0$.

Starting from raw data, we first subtract the constant signal at
$\omega >0$ (due to second order light). Next an `unoccupied' state
spectrum  at a $k$ far from $k_F$ is used as an energy-dependent
background \cite{ADAMBG}.
The subtraction is performed by normalizing the background to each
spectra at a given binding energy, $\omega_c$ (320 meV for the data
set in question), and then subtracting it \cite{BG}.
This effective spectral function represents the
the renormalized band near the Fermi energy.
Finally, we divide the data by a resolution \cite{RESOLUTION} broadened
Fermi function to obtain ${\rm Im}G({\bf k},\omega)$ for $\omega < 0$.

The next step is to determine the unoccupied part of the spectral
function, ${\rm Im}G({\bf k},\omega)$ for $\omega > 0$, which cannot
be obtained directly from ARPES data. We obtain this by invoking
particle-hole symmetry with respect to the Fermi surface: ${\bf
k}_F$, Im$G({\bf k}_F+{\bf k},\omega)={\rm Im}G({\bf k}_F-{\bf
k},-\omega)$, where ${\bf k}$ is directed along the normal to the
Fermi surface.  This assumption should be reasonable in the
superconducting state of optimally doped cuprates over an energy
range in excess of the gap, as evidenced by the approximate
particle-hole symmetry seen in tunneling experiments \cite{STM}. We
have also checked that this assumption does not qualitatively affect
our final results for $\chi_0$ (by using ${\rm Im}G$ with p-h
asymmetry put in by hand). We then normalize the obtained ${\rm
Im}G$ so that the integral of the spectral function ($-{\rm
Im}G/\pi$) is equal to unity over the energy range of $\pm
\omega_c$.  This minimizes the effect of dipole matrix elements. Now
we may use the ${\rm Im}G$ derived from ARPES to calculate
$\chi_0^G$ (we will discuss $\chi_0^F$ later).
Finally, to perform the ${\bf k}$-sum in Eq.~(\ref{eq:chi0}), the
ARPES data are interpolated to a regular grid and then reflected
using square lattice group operations to fill the first Brillouin
zone \cite{UTPAL}. We used a $100 \times 100$ grid.

\begin{figure}[!h]
\includegraphics[scale=0.45]{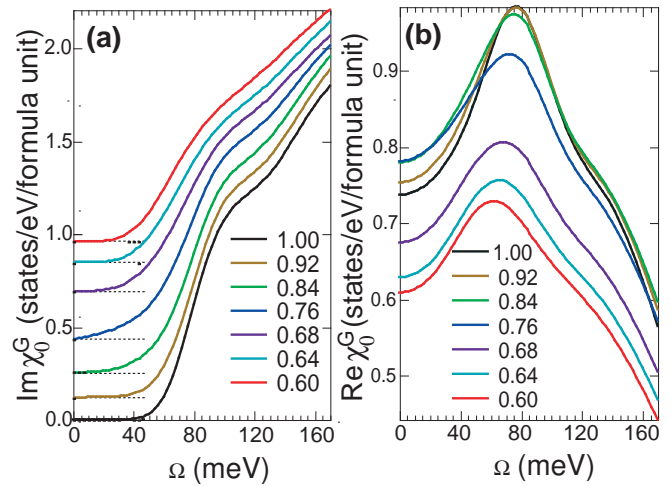}
\caption{(a) Im$\chi^G_{0}$ and (b) Re$\chi^G_0$ in the
superconducting state as a function of frequency for momenta along
$\bf{q}=\eta (\pi,\pi)$. The various curves, labeled by $\eta$,
are offset for clarity in panel (a), noting that Im$\chi^G_{0}=0$ at $\Omega=0$.}
\label{fig:Chiq}
\end{figure}

Fig.~1a shows the calculated Im$\chi^G_0$ at T=40K
(superconducting state)  as a function of the momentum transfer
${\bf q}$ along the zone diagonal. We note that Im$\chi^G_0$ is
greatly suppressed at low energies due to the gap to particle-hole
(p-h) excitations in the superconducting state and then increases
quite sharply. The p-h gap is in general given by the sum of the
superconducting gaps at two points on the Fermi surface separated by
the wavevector ${\bf q}$. The ${\bf Q}=(\pi,\pi)$ vector connects
the hot spots ($\epsilon_{\bf k}=\epsilon_{\bf k+Q}$=0) which are
not too far from the zone boundary in Bi2212, and thus the hot-spot
gap is comparable to the one at the antinode. Consequentially at
${\bf Q}=(\pi,\pi)$, we see in Fig.~1a a large gap whose midpoint
is around 80 meV, roughly twice the maximum d-wave superconducting
gap of $\simeq$ 40 meV at the antinode.
We note that the threshold is quite broad ($\sim$ 40 meV) as a
result of the intrinsic broadening of Im$G$ arising from
self-energy effects as well as resolution. As $q$ decreases from
$Q$, one sees the p-h gap decrease due to the d-wave anisotropy of
the gap, and then disappear at ${\bf q}_n \simeq (0.76,0.76)\pi$.
${\bf q}_n$ is the wavevector corresponding to node-node scattering
with the d-wave gap vanishing at the nodes. For $q < q_n$, the p-h
gap reappears \cite{RPA}.

In Fig.~1b we show Re$\chi^G_0$ obtained from Eq.~(\ref{eq:chi0})
\cite{DELTA}.   First concentrating on ${\bf Q}$, we note the
presence of a peak that corresponds to the gap midpoint of Fig.~1a
as expected from Kramers-Kronig relations. This peak is broadest for
${\bf q}_n$ where the p-h gap vanishes in the imaginary part.

We now turn to $\chi^F_0$. Since ${\rm Im} F$ is not available from
experiment, we estimate the FF term as follows.  We calculate the
BCS $\chi^G_{0,{\rm BCS}}$ and $\chi^F_{0,{\rm BCS}}$ from
Eq.~(\ref{eq:chi0})  using the bare BCS Green's functions
$G_{0}({\bf k},\omega)= (\omega + \epsilon_{\bf k})/(\omega^2 -
\epsilon^2_{\bf k} - \Delta_{\bf k}^2)$ and $F_{0}({\bf k}, \omega)=
\Delta_{\bf k}/(\omega^2 - \epsilon^2_{\bf k}- \Delta_{\bf k}^2)$
with the experimentally measured dispersion \cite{DISPERSION}
$\epsilon_{\bf k}$ and the measured $\Delta_{\bf k}$, which we find
to be proportional to $(\cos k_x-\cos k_y)$ for this sample.
We define the ratio of the real and imaginary parts, given by
$\alpha_R({\bf q},\Omega)= {\rm Re} \chi^F_{0,{\rm BCS}} / {\rm Re}
\chi^G_{0,{\rm BCS}}$ and similarly for $\alpha_I({\bf q},\Omega)$.
We then assume that the missing contribution $\chi_0^F$  may be
accounted for by
${\rm Re} \chi_0={\rm Re}\chi_0^G + {\rm Re}\chi_0^F  \simeq
(1+\alpha_R) {\rm Re}\chi^G_0$ and ${\rm Im} \chi_0={\rm Im}\chi_0^G
+ {\rm Im}\chi_0^F \simeq (1+\alpha_I) {\rm Im}\chi^G_0$. We will
discuss below the extent to which our final results are affected by
this approximation.

In order to carry out comparisons with INS data, or other probes
such as NMR, the full spin susceptibility $\chi$ is needed. The most
common approximation is to use the RPA form \cite{RPA}
\begin{equation}
\chi({\bf q},\Omega)=\frac{\chi_{0}({\bf q},\Omega)}{1-U({\bf q})
\chi_{0}({\bf q},\Omega)} \label{eq:RPA}
\end{equation}
where $U$ is an effective interaction.  In this paper, we will assume two limiting
forms for this effective interaction,  one where it is a constant ($U_0$), the other where it has
the form
$U({\bf q})= - U_0(\cos q_x +\cos q_y )/2$ corresponding to
superexchange between near neighbor copper sites.

The experimental results for $\chi_0$ presented in
Fig.~1 are similar to those obtained from BCS theory \cite{RPA}
especially at low energies where the incoherent spectral weight in
the ARPES data is small. Within BCS theory, which uses \emph{bare}
Green's functions, one has a true gap in Im$\chi_{0}$ at ${\bf
Q}$ and a corresponding log divergence in Re$\chi_{0}$. These
features still persist in Fig.~1 albeit broadened by self-energy
(and resolution) effects. As such, for some frequency smaller than
the gap, one will obtain a pole in $\chi$ when $1-U({\bf q}){\rm Re}
\chi_0 ({\bf q}, \Omega)=0$ provided Im$\chi_0$ is small
 at the frequency of interest. This pole represents a
collective mode, known as the spin resonance at ${\bf Q}$, which is
prominently observed in INS data for YBCO \cite{NEUTRON} and Bi2212
\cite{INSB,FAUQUE}. Following this logic, we fix $U_0$ \cite{U0} by
fitting the energy (44 meV) of the INS spin resonance at ${\bf Q}$
for optimal doped Bi2212 \cite{INSB}.

\begin{figure}[!h]
\includegraphics[scale=0.45]{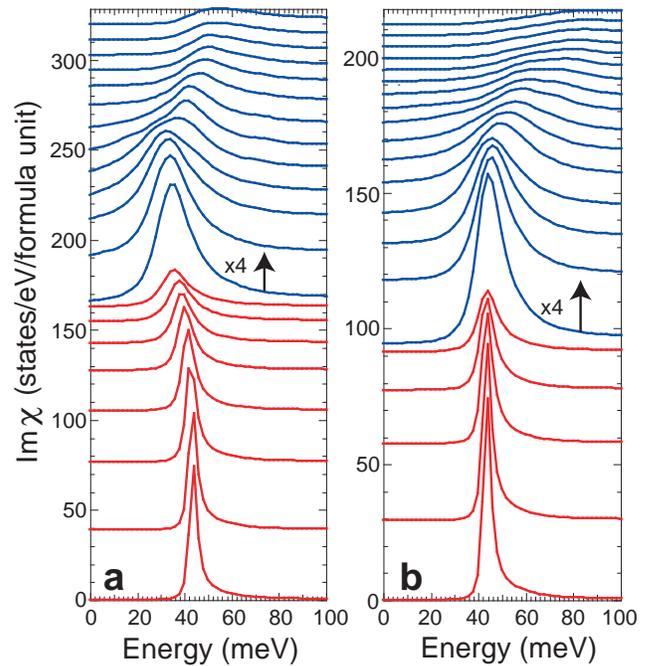}
\caption{Im$\chi$ as a function of energy for several momenta ${\bf
q}=\eta(\pi,\pi)$ assuming (a) a constant $U=U_0$ or (b) a near
neighbor exchange $U=U({\bf q})$. The curves,  ranging from
$\eta$=0.60 (top) to $\eta$=1 (bottom), are offset for clarity. Some
curves are rescaled for visual clarity.}
\label{fig:respeak}
\end{figure}

\begin{figure}[!h]
\includegraphics[scale=0.45]{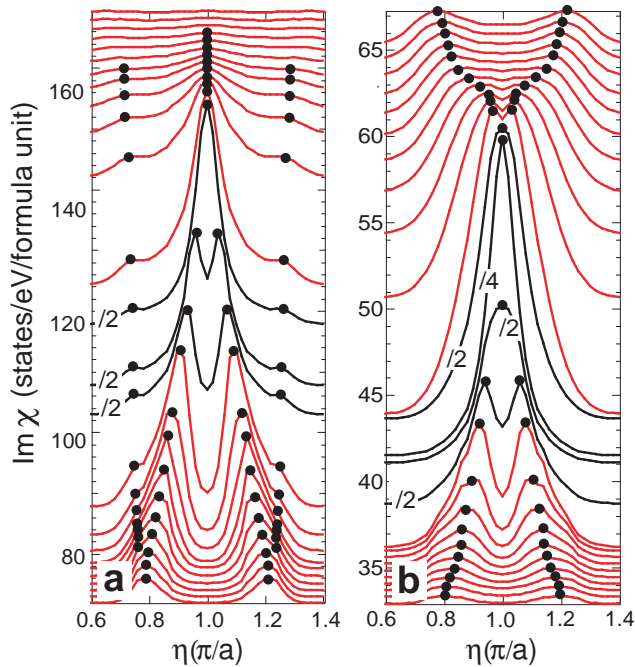}
\caption{Im$\chi$ as a function of momentum ${\bf q}=\eta(\pi,\pi)$
for several energies assuming (a)  a constant $U=U_0$ or (b) a near
neighbor exchange $U=U({\bf q})$. The curves, ranging from
$\Omega$=68 meV (top) to 20 meV (bottom), are offset for clarity.
Some curves are rescaled for visual clarity.}
\label{fig:res_disp}
\end{figure}

In order to compare the results of our approach with INS data we
plot the imaginary part of the full susceptibility $\chi$ as
calculated from ARPES data as discussed above for constant ${\bf q}$
cuts in Fig.~\ref{fig:respeak} and constant $\Omega$ cuts in
Fig.~\ref{fig:res_disp}. The left panels assume a constant $U$, the
right panels the near neighbor exchange form $U({\bf q})$. Let us
first consider Fig.~\ref{fig:respeak}. For constant $U$ (left panel), the resonance
traces out a pronounced downwards dispersion which terminates at the
node-node scattering vector ${\bf q}_n$, as seen in the INS data.
For $q < q_n$, a distinct second branch appears, which is broad and
weak, that disperses upwards.  The change of behavior at ${\bf q}_n$
corresponds to the so-called silent band effect and second mode
mentioned in connection with INS data of overdoped YBCO
\cite{PAILHES,EREMIN} and Bi2212 \cite{FAUQUE}. But interestingly,
the upper branch of the `hourglass' is not apparent. In contrast,
for $U({\bf q})$ (right panel), the mode is almost dispersionless
near $(\pi,\pi)$, then shows an {\it upward} dispersion for momenta ${\bf
q}<0.9 (\pi,\pi)$. This difference is a
direct consequence of the relatively weak momentum dependence of
${\rm Re}\chi_0$ (Fig.~1b) coupled to the decrease in $U({\bf q})$
away from $Q=(\pi,\pi)$.

The behavior of the dispersion observed in constant ${\bf q}$ scans can
be contrasted with that from constant $\Omega$ scans
(Fig.~\ref{fig:res_disp}). For constant $U$, both types of scans
yield a qualitatively similar mode dispersion (solid dots in
Fig. ~\ref{fig:res_disp}a). However, for $U({\bf q})$, the
dispersion obtained from the constant $\Omega$ cuts is hourglass-like,
with an upward and downward branch that merge at $(\pi,\pi)$, in
good agreement with INS data in underdoped YBCO \cite{HAYDEN}.  We note
that this downward branch is not visible in the constant
${\bf q}$ cuts in Fig.~\ref{fig:respeak}b.
This difference is analogous to the different ARPES dispersions that one finds
from energy distribution curves as compared to momentum distribution curves.

We have also generated results involving only the GG contribution by
setting $\alpha= 0$. In order to compare with the results  having
both FF and GG contribution, we rescale $U_0$ to maintain the same
resonance energy at ${\bf Q}$.  Only minor differences are found
between this and the full calculation which includes both GG and FF
contributions. This lack of sensitivity to the inclusion of FF terms
is a consequence of the d-wave symmetry of the superconducting order
parameter where $\chi^G_0$ and $\chi^F_0$ reinforce one another near
$(\pi,\pi)$.  (In contrast, for an s-wave superconductor, $\chi^G_0$
and $\chi^F_0$ are opposite in sign, and there is no spin
resonance.)

Comparing our results with the INS data and earlier RPA
calculations, we arrive at the following conclusions. First, the
self-energy effects present in the ARPES-derived Green's function do
not affect the low energy physics of spin excitations, such as the existence and sharpness of
the $(\pi,\pi)$ resonance, or the mode dispersion. Second, vertex
corrections do not play a major role in the spin channel, except possibly in
the overall scale of $U$.
Third, the magnetic dispersion is sensitive to the ${\bf q}$ dependence of $U$.
Finally, our results provide strong evidence for the
interpretation of the resonance peak as a spin exciton \cite{RPA}.

To summarize, we computed the polarization bubble $\chi_0$ using
experimental Green's functions derived from ARPES spectra, and the
full dynamic spin response obtained from a diagrammatic formalism
assuming either a constant or a near neighbor exchange interaction. Although this
analysis requires several approximations, we find surprisingly good
agreement with inelastic neutron scattering data for high
temperature cuprate superconductors. Our results demonstrate a close
relation between experiments probing the spin and single-particle
excitations. Our formalism is quite general, and can be used as well
to compute other response functions, such as the current-current
response function measured by conductivity.

This work was supported by NSF DMR-0606255, NSF DMR-0513415 (D.K.M.)
the U.S. DOE, Office of Science, under Contracts No.~DE-AC02-06CH11357
(ANL), W-7405-Eng-82 (Ames), and Grant No. DE-FG02-05ER46225
(D.K.M.). The Synchrotron Radiation Center is
supported by NSF DMR-0084402. U.C. would like to thank R. Sensarma
for helpful discussions.


\begin{thebibliography}{99}

\bibitem{REVIEW}
A. Damascelli, Z. Hussain and Z.-X. Shen, Rev. Mod. Phys. {\bf 75}, 473 (2003);
J. C. Campuzano, M. R. Norman and M. Randeria, in {\it The Physics of Superconductors}, Vol. 2, ed. K. H. Bennemann and J. B. Ketterson (Springer, Berlin, 2004), p. 167.

\bibitem{NEUTRON}
J. Rossat-Mignod \etal, Physica C {\bf 185}, 86 (1991);
H. F. Fong \etal, Phys. Rev. Lett. {\bf 75}, 316 (1995);
P. Dai  \etal, Phys. Rev. B {\bf 63}, 054525 (2001).

\bibitem{RPA} J. Brinckmann and P. A. Lee, Phys. Rev. Lett. {\bf 82}, 2915 (1999);
Y.-J. Kao, Q. Si and K. Levin, Phys. Rev. B {\bf 61}, 11898 (2000);
M. R. Norman, Phys. Rev. B {\bf 61}, 14751 (2000) and {\bf 63}, 092509 (2001);
F. Onufrieva and P. Pfeuty, Phys. Rev. B {\bf 65}, 054515 (2002).

\bibitem{OTHERS}
See, e.g.,
C.-H. Pao and N. E. Bickers, Phys. Rev. B {\bf 51}, 16310 (1994);
E. Demler and S.-C. Zhang, Phys. Rev. Lett. {\bf 74}, 4126 (1995);
N. Bulut and D. Scalapino, Phys. Rev. B {\bf 53}, 5149 (1996).

\bibitem{UHRIG}
M. Vojta and T. Ulbright, Phys. Rev. Lett. {\bf 93}, 127002 (2004);
G. S. Uhrig, K. P. Schmidt and M. Grunninger, Phys. Rev. Lett. {\bf 93}, 267003 (2004);
C. D. Batista, G. Ortiz and A. V. Balatsky, Phys. Rev. B {\bf 64}, 172508 (2001).

\bibitem{INSB}
H. F. Fong \etal, Nature {\bf 398}, 588 (1999).

\bibitem{ARAI}
M. Arai \etal, Phys. Rev. Lett. {\bf 83}, 608 (1999).

\bibitem{HAYDEN}
S. M. Hayden \etal, Nature {\bf 429}, 531 (2004).

\bibitem{FAUQUE}
B. Fauque \etal, cond-mat/0701052.

\bibitem{KINK}
A. Kaminski \etal, Phys. Rev. Lett. {\bf 86}, 1070 (2001).

\bibitem{UTPAL}
U.Chatterjee {\it et al.}, Phys. Rev. Lett. {\bf 96}, 107006 (2006).

\bibitem{MAHAN}
G. D. Mahan, {\it Many-Particle Physics} (Plenum, New York, 1990).

\bibitem{BOB}
J. R. Schrieffer, {\it Theory of Superconductivity} (Benjamin Cummings, Reading, MA, 1964).

\bibitem{NK}
M. Randeria {\it et al.}, Phys. Rev. Lett. {\bf 74}, 4951 (1995).

\bibitem{ADAMBG}
A. Kaminski {\it et al.}, Phys. Rev. B {\bf 69}, 212509 (2004).

\bibitem{BG}
The background is thought to be due to surface scattering.  We have checked that our final results for
$\chi$  are not qualitatively affected by its subtraction.

\bibitem{RESOLUTION}
ARPES data are also affected by finite energy and momentum resolutions which,
in principle, can be deconvolved from the data but, in practice, have been found
not to have a significant effect on Im$G$. See: M. R. Norman \etal, Phys. Rev. B {\bf 60}, 7585 (1999).

\bibitem{STM} See, e.g., N. Miyakawa \etal, Phys. Rev. Lett. {\bf 83}, 1018 (1999);
S. H. Pan \etal, Nature {\bf 413}, 282 (2001).

\bibitem{DELTA} We take a finite $\delta = 0.5$ meV for the calculation of Re$\chi^G_0$.

\bibitem{DISPERSION}
M. R. Norman \etal, Phys. Rev. B {\bf 52}, 615 (1995).

\bibitem{U0}
The resulting value of $U_0$ is 924 meV, a reasonable number  for an
effective U.

\bibitem{PAILHES}
S. Pailhes \etal, Phys. Rev. Lett. {\bf 93}, 167001 (2004).

\bibitem{EREMIN}
I. Eremin \etal, Phys. Rev. Lett. {\bf 94}, 147001 (2005).

\end{thebibliography}
\end{document}